\documentclass[aps,prd,showpacs,twocolumn,standalone,10pt,superscriptaddress,floatfix,nofootinbib,longbibliography]{revtex4-2}
\usepackage[T1]{fontenc}
\usepackage[utf8]{inputenc}
\usepackage{amssymb,amsfonts,amsmath,bm}
\usepackage{graphicx}
\usepackage{dcolumn}
\usepackage{booktabs}
\usepackage{multirow}
\usepackage{makecell}

\usepackage[
colorlinks=true,
linkcolor=blue,
breaklinks=true,
urlcolor=blue,
citecolor=blue]{hyperref}

\graphicspath{{figures/}}

\begin{document}

\title{Adaptive Matrix-Based Denoising Algorithm Using Local Spatial Support}

\author{Zhaoli Guo}
\affiliation{School of Physics and Electronic Science, Hunan University of Science and Technology, Xiangtan 411201, China}
\author{Qingyu Li}
\affiliation{School of Physics and Electronic Science, Hunan University of Science and Technology, Xiangtan 411201, China}
\author{Baolin Zhang}
\affiliation{School of Physics and Electronic Science, Hunan University of Science and Technology, Xiangtan 411201, China}
\author{Jiarui Zhao}
\affiliation{School of Physics and Electronic Science, Hunan University of Science and Technology, Xiangtan 411201, China}
\author{Aonan Zhu}
\affiliation{School of Physics and Electronic Science, Hunan University of Science and Technology, Xiangtan 411201, China}
\affiliation{Hunan Provincial Key Laboratory of Intelligent Sensors and New Sensor Materials, School of Physics and Electronic Science, Hunan University of Science and Technology, Xiangtan 411201, China}
\author{Liang Peng}\email[First corresponding author:~]{pengliang@hnust.edu.cn}
\affiliation{School of Physics and Electronic Science, Hunan University of Science and Technology, Xiangtan 411201, China}
\affiliation{Hunan Provincial Key Laboratory of Intelligent Sensors and New Sensor Materials, School of Physics and Electronic Science, Hunan University of Science and Technology, Xiangtan 411201, China}
\author{Huilin Li}\email[Second corresponding author:~]{lihuilin@hnust.edu.cn}
\affiliation{School of Physics and Electronic Science, Hunan University of Science and Technology, Xiangtan 411201, China}
\affiliation{Hunan Provincial Key Laboratory of Intelligent Sensors and New Sensor Materials, School of Physics and Electronic Science, Hunan University of Science and Technology, Xiangtan 411201, China}

\begin{abstract}
To address the online data-processing requirements of the high-level trigger (HLT) under high-event-rate and high-background conditions in high-luminosity collider facilities, this paper proposes an adaptive matrix-based denoising algorithm using local spatial support. During online operation, the algorithm requires neither parameter scanning nor neural-network inference. Instead, an occupancy-dependent adaptive-threshold selection rule is obtained through offline calibration, thereby improving the adaptability to different background conditions while maintaining low online computational overhead. The algorithm is implemented and tested in the main drift chamber (MDC) of the Super Tau-Charm Facility (STCF). By mapping each superlayer to a two-dimensional binary matrix, isolated hits and small-scale point-cluster backgrounds with insufficient local spatial support are conservatively filtered through sliding-window accumulation, local-neighborhood enhancement, and thresholding. To accommodate different event occupancies and degrees of local clustering, the global occupancy and window activation ratio are used as inputs, and an event-level adaptive threshold is determined using an offline-calibrated low-order polynomial. Based on tests with five charged-final-state physics channels simulated using the Offline Software of the Super Tau-Charm Facility (OSCAR), the rejection rate for discrete point-cluster backgrounds exceeds $90\%$ at low to moderate background levels, while the true-hit retention ratio remains above $98\%$. Even when the detector efficiency is reduced to $90\%$, the true-hit retention ratio remains above $98\%$. In Hough track-finding tests at one nominal background level, the proposed denoising algorithm reduces the average track-finding time per event from approximately $116~\mathrm{ms}$ to approximately $96~\mathrm{ms}$. On a single GPU, the algorithm achieves a theoretical throughput of approximately $2.7\times10^{6}~\mathrm{events/s}$. These results demonstrate that the proposed algorithm satisfies the stringent online-processing constraints of an HLT upstream denoising module, including low and predictable average processing time, high precision, and high throughput. Because its filtering criterion is derived from local spatial relationships among detector hits rather than predefined track parameters or a model of a specific physical process, the proposed algorithm provides a general research approach for online denoising in regularly structured detectors used in high-luminosity collider facilities.
\end{abstract}


\maketitle

\section{Introduction}

In high-luminosity particle collider experiments, detectors must continuously acquire data under high event rates and complex background conditions. As the number of collision events per unit time increases, the volume of raw data produced by the detector also grows continuously. Background hits further increase the computational burden associated with data transmission, buffering, online reconstruction, and offline analysis. Therefore, appropriately filtering raw detector hits online while preserving genuine physics signals as completely as possible is an important issue in data processing for high-luminosity experiments.

The Super Tau-Charm Facility (STCF) is a next-generation large-scale scientific facility currently planned in China\cite{STCFdesign}. It is designed to operate in the tau-charm energy region, with a center-of-mass energy range of $2$--$7~\mathrm{GeV}$. This energy region covers important threshold domains for light hadrons, charmed hadrons, and tau leptons, and provides an important window for studying non-perturbative effects in quantum chromodynamics, performing precision tests of the Standard Model, and searching for new physics beyond the Standard Model. While offering substantial physics potential, the STCF also imposes more stringent requirements on its detector, trigger, and data-acquisition systems. According to the current design of the STCF trigger system, the peak physics event rate is approximately $400~\mathrm{kHz}$, and the raw data throughput is approximately $30~\mathrm{GB/s}$. High-luminosity operation means that the number of collision events occurring per unit time increases significantly, and the detector must operate stably at a higher event rate.

To reduce the pressure on data storage, the STCF is designed with a high-level trigger (HLT) system\cite{STCFdesign,ATLASTDAQ,ATLASHLT,ATLASHLT2,CMSHLT}. The HLT is located after the Level-1 (L1) trigger and the event-building stages of the data-acquisition system, but before data storage. Its primary task is to filter out background hits from the same time period during online processing, thereby reducing the pressure on data storage and improving the efficiency of subsequent data processing. The HLT currently faces processing pressure caused by detector background hits. Taking the main drift chamber (MDC) as an example, the MDC is a cylindrical multi-wire drift chamber, which is a gas detector that detects charged particles through the ionization they produce in the gas. Drift chambers offer a low material budget, low cost, and good position resolution, and are therefore widely used in particle-physics experiments such as BESIII and Belle~II. Under the high-luminosity operating conditions of the STCF, raw MDC hit data are affected by complex background processes. These backgrounds can generally be divided into beam-related backgrounds and luminosity-related backgrounds. Beam-related backgrounds mainly include the Touschek effect and beam--gas interactions, including Coulomb scattering and bremsstrahlung, while luminosity-related backgrounds mainly include radiative Bhabha scattering and two-photon processes. Because the MDC responds to charged particles traversing its sensitive volume, both background particles and genuine charged particles produced by the target physics processes generate hits on the sense wires. These background hits not only consume additional resources and increase the processing time of reconstruction algorithms in the HLT, but may also reduce their reconstruction efficiency. Without effective filtering and removal at the initial stage of the HLT, background hits would increase the pressure on intermediate data transfer and buffering within the HLT, place an additional computational burden on subsequent online reconstruction algorithms, and degrade reconstruction performance. Therefore, it is important to develop a denoising algorithm that can rapidly process raw detector hit data, effectively remove background hits, and preserve genuine physics hits as completely as possible.

In recent years, neural networks and other learning-based methods have been applied to detector-hit identification and background suppression\cite{JiaGNN2026,ThomadakisDenoising2022}. These methods can exploit hit positions, drift times, charges, and geometric correlations across detector layers to achieve good filtering performance on specific datasets. However, they typically rely on relatively complex feature construction, model inference, or global information association. For an HLT upstream denoising module, filtering performance is not the only consideration; processing latency, resource consumption, parallel efficiency, and the risk of erroneously rejecting genuine hits must also be taken into account. Therefore, upstream denoising in the HLT should not be formulated as a complete background-classification or track-recognition problem. Instead, online denoising methods based on the physical characteristics of the detector, with regular computational procedures and low processing latency, should be investigated.

Spatial filtering based on the local spatial regularity of physics signals provides a promising approach to this problem. As charged particles traverse a drift chamber, they typically leave hits with a certain degree of spatial continuity in adjacent or neighboring detector cells. In contrast, some beam-related and luminosity-related backgrounds are more likely to appear as isolated hits or small hit clusters with insufficient local spatial support. By enhancing local spatial correlations among hits and suppressing hits lacking neighborhood support, a portion of clearly irrelevant background input can be removed at the raw-hit level, thereby alleviating the computational pressure on subsequent track-finding and reconstruction algorithms. Because backgrounds do not necessarily exhibit completely random patterns, some background hits may also form local structures similar to genuine tracks. Consequently, upstream spatial filtering is more appropriately implemented using a conservative strategy: it should focus on hits with clearly insufficient local support, while retaining hits whose classification may require cross-layer associations or global reconstruction information.

This paper proposes an adaptive matrix-based denoising algorithm using local spatial support. The algorithm represents the hit information in a detector superlayer as a two-dimensional binary matrix, performs sliding-window statistics to evaluate the degree of local hit support, enhances local structures using neighborhood relationships, and finally retains hits with sufficient spatial support according to a threshold. The algorithm does not depend on complete track parameters, cross-superlayer pattern recognition, or global track fitting. Instead, it mainly exploits the local spatial organization of detector hits to perform the filtering. On this basis, an adaptive threshold-selection mechanism based on the event occupancy and the local window activation ratio is further introduced. The corresponding mapping relationship is calibrated using offline simulation samples. During online operation, the filtering threshold is calculated using a fixed low-order function according to the hit occupancy and local window status of the current event. Therefore, the algorithm does not require the online scanning of a large number of candidate parameters or complex model inference, and can adapt to different background levels and event occupancies within a certain range. The three main processing steps of the algorithm---sliding-window accumulation, local-neighborhood enhancement, and threshold filtering---all have regular data-access patterns and a high degree of data parallelism. Its computational procedure is therefore relatively stable and suitable for batch processing on GPUs, allowing a balance to be achieved among processing throughput, computational resource consumption, and physics-filtering performance.

The proposed matrix-based denoising algorithm is validated in the HLT environment of the STCF MDC using OSCAR simulation data for five charged-final-state physics channels at different background levels. At low to moderate background levels, the rejection rate for discrete point-cluster backgrounds exceeds $90\%$, while the true-hit retention ratio remains above $98\%$. Even when the detector efficiency is reduced to $90\%$, the true-hit retention ratio remains above $98\%$. When used as an upstream denoising module for Hough-transform track finding, the algorithm reduces the average track-finding time per event from approximately $116~\mathrm{ms}$ to approximately $96~\mathrm{ms}$ at one nominal background level. On a Tesla A16 GPU, the denoising algorithm requires only approximately $14.6\%$ of the computational capacity of one GPU at an event rate of $400~\mathrm{kHz}$ and achieves a theoretical throughput of approximately $2.7\times10^{6}~\mathrm{events/s}$. These results demonstrate that the proposed algorithm satisfies the stringent online-processing constraints of an HLT upstream denoising module, including low and predictable average processing time, high precision, and high throughput. Because its filtering criterion is derived from local spatial relationships among detector hits rather than predefined track parameters or a model of a specific physical process, the proposed algorithm provides a general research approach for online denoising in regularly structured detectors used in high-luminosity collider facilities.

\section{MDC and HLT Design}

\subsection{Main Drift Chamber}

The main drift chamber (MDC) is an important component of the tracking system of the STCF detector. It is designed as a cylindrical multi-wire drift chamber. When charged particles pass through the working gas, they ionize gas molecules, and the resulting ionization electrons drift toward the sense wires under the electric field, producing readable electrical signals. The spatial trajectories of charged particles can be reconstructed from hit information recorded in multiple detector layers, from which physical quantities such as momentum can be further obtained. The MDC contains 48 layers of drift cells, with every six layers forming one superlayer. All layers within the same superlayer contain the same number of cells. The entire tracking system is placed in a uniform magnetic field of $1~\mathrm{T}$ along the $z$ axis, and the polar-angle coverage is $20^\circ$--$160^\circ$. The MDC system contains a total of 11,520 readout cells. Several key design parameters of the MDC are listed in Table~\ref{tab:mdc_parameters}. The MDC contains field wires and sense wires. The sense wires collect the signals produced by the ionization of charged particles and are mainly divided into axial wires and stereo wires. The axial wires are parallel to the beam direction, whereas the stereo wires are arranged at a certain angle with respect to the beam direction. Information such as the time and charge measured from these signals provides the basis for subsequent track reconstruction and particle identification.

Under the high-luminosity operating conditions of the STCF, raw MDC hit data are affected by complex background processes. These backgrounds can generally be divided into beam-related backgrounds and luminosity-related backgrounds. Beam-related backgrounds mainly include the Touschek effect and beam--gas interactions, including Coulomb scattering and bremsstrahlung, whereas luminosity-related backgrounds mainly include radiative Bhabha scattering and two-photon processes\cite{OSCAR3,STCFBackground}. Because the MDC responds to charged particles traversing its sensitive volume, both background particles and genuine charged particles produced by the target physics processes generate hits on the sense wires.

In a gas drift chamber, the formation of a valid hit is jointly affected by particle ionization, gas amplification, electronic readout, and other stages. Because primary ionization and signal amplitudes fluctuate statistically, some signals may fall below the electronics threshold. In addition, if the drift time falls outside the effective readout window, the signal may not be recorded as a valid hit. Dead channels, locally inefficient regions, and geometric boundary effects can also reduce the local hit efficiency. Under high-luminosity operating conditions, beam-related backgrounds increase the channel occupancy and can consequently cause signal pileup and electronic dead time, leading to further loss of valid hits.

The background processes and detector-efficiency effects described above result in raw MDC hit data containing both additional background hits and local gaps in genuine tracks. Both effects must be considered when designing MDC algorithms for the HLT. In particular, during upstream denoising, reduced detector efficiency may cause genuine tracks to appear as incomplete short track segments or locally isolated hits. While removing isolated background hits and small noise clusters, the denoising algorithm should avoid misclassifying isolated hits caused by missing portions of genuine tracks as noise.
\begin{table}[t]
	\centering
	\caption{The main geometric parameters of the Main Drift Chamber}
	\label{tab:mdc_parameters}
	\begin{tabular}{ccccc}
		\hline
		\textbf{Type} & \textbf{Radius (mm)} & \textbf{Layers} & \textbf{Units/Layer} & \textbf{Stereo Angle (mrad)} \\
		\hline
		A & 200.0 & 6 & 128 & 0 \\
		U & 271.6 & 6 & 160 & 39.3 to 47.6 \\
		V & 342.2 & 6 & 192 & -41.2 to -48.4 \\
		A & 419.2 & 6 & 224 & 0 \\
		U & 499.8 & 6 & 256 & 50.0 to 56.4 \\
		V & 578.1 & 6 & 288 & -51.3 to -57.2 \\
		A & 622.0 & 6 & 320 & 0 \\
		A & 744.0 & 6 & 352 & 0 \\
		\hline
		Total & & 48 & 11520 & \\
		\hline
	\end{tabular}
\end{table}

\subsection{HLT Design and Requirements for the Denoising Algorithm}
The STCF trigger system includes a Level-1 (L1) hardware trigger\cite{STCFL1trigger,STCFL1trigger2,HuangL1STCF2026}, followed by the HLT, where more complete event information is available for online processing. Online processing of MDC background hits can generally be understood from two perspectives: directly removing disordered background hits that lack physical associations, or identifying and retaining signal hits with ordered track-like structures. The latter is the more natural approach in complete track finding and reconstruction. The trajectories produced by charged particles in the MDC, as well as the local track segments formed in individual superlayers, have distinct spatial structures. Pattern matching, global track finding, and track fitting can preserve the physics signals of interest by identifying these ordered structures and can suppress background hits indirectly in the process\cite{HuangL1STCF2026,JiaSegmentHLT2026,ZhouHough2025,RohrALICEGPU2018,BaileyBABAR2004,LantzParallelKalman2020,ChangSegmentLinking2022}.

However, this approach is not suitable as the primary strategy for MDC denoising at the front end of the HLT. Upstream MDC denoising in the HLT takes place before track finding, track segment finding and track segment association, and track fitting. It is essentially a local hit-level filtering problem subject to an extremely short processing-time constraint, rather than a complete track-finding or background-classification problem. If all known signals were identified at this stage and all remaining hits were removed, the algorithm would need to reliably recognize genuine track structures across different background levels, physics channels, and momentum ranges. Once genuine signals are mistakenly removed at the upstream stage, subsequent track-finding and reconstruction algorithms have little chance of recovering the lost information. Therefore, upstream denoising must prioritize the retention of genuine hits rather than pursue the complete removal of all background hits.

\begin{figure}[!htb]
	\centering
	\includegraphics[width=0.95\columnwidth]{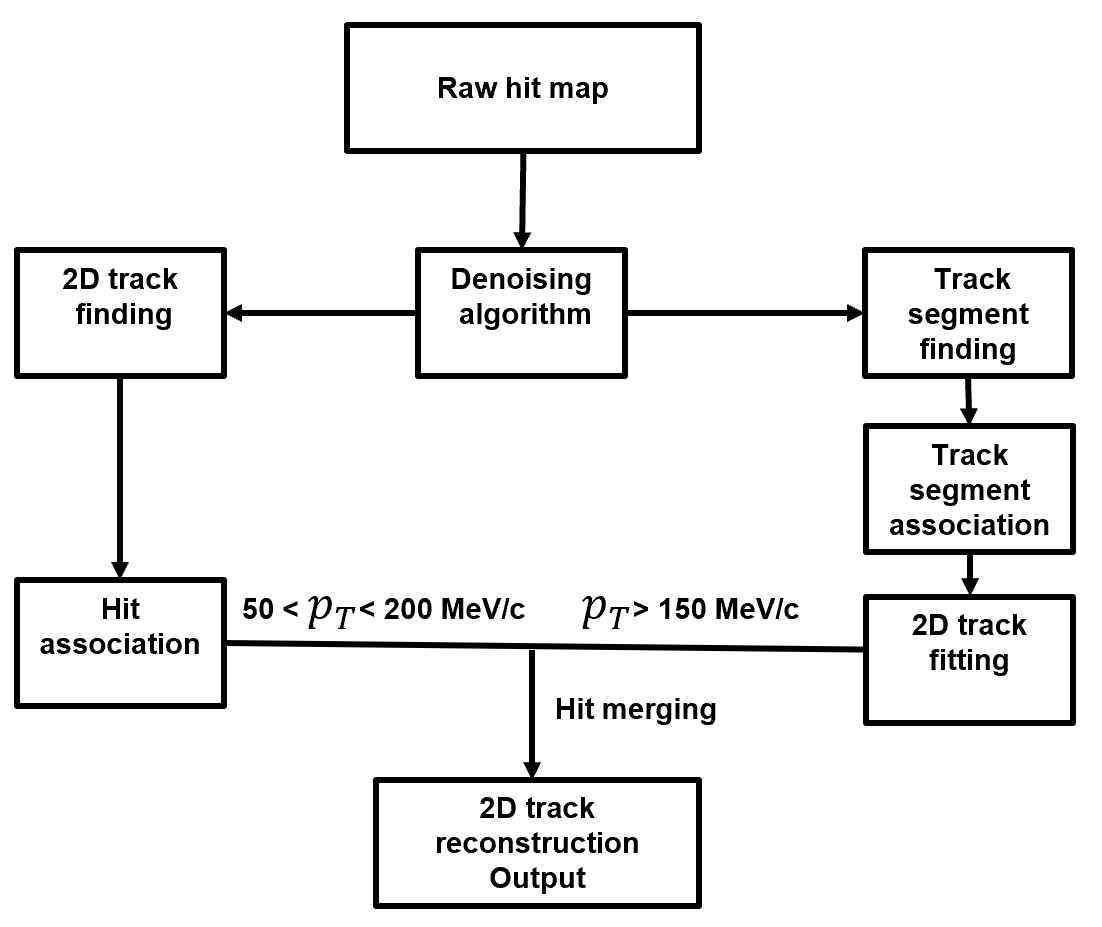}
	\caption{Design of the two-dimensional track-finding and reconstruction workflow for the MDC at the HLT stage. The hit information produced by the denoising algorithm is distributed to two branches. The left branch is intended to use a Hough-transform algorithm to reconstruct tracks in the lower-transverse-momentum region. The right branch uses pattern matching and non-recursive depth-first search (DFS) to identify candidate track segments and their associations, followed by linear fitting to extract the track parameters. During the merging stage, each branch outputs only the tracks and their associated hits within its assigned transverse-momentum range. A $50~\mathrm{MeV}/c$ overlap margin is reserved between the transverse-momentum coverage ranges of the two branches.}
	\label{fig:mdc_hlt_2d_reconstruction}
\end{figure}

The background hits in MDC data are also not simply completely random isolated points or small clusters. In addition to isolated points and small hit clusters, some background hits have local spatial distributions very similar to those of genuine particle tracks. Although these hits are labeled as background in simulation samples, they may have shapes within a single superlayer that resemble genuine track segments. In some cases, background hits may even form structures resembling charged-particle tracks across the full MDC. Judging such hits using only local two-dimensional information during online processing can easily lead to erroneous removal. For hits whose physical origin can be determined only through global associations, timing information, fit residuals, or offline analysis, the upstream denoising stage should not remove them indiscriminately. This consideration is also related to the overall HLT algorithm design. Fig.~\ref{fig:mdc_hlt_2d_reconstruction} illustrates the current design of the MDC two-dimensional track-finding and reconstruction workflow at the HLT stage. In the MDC HLT chain, subsequent two-dimensional track finding and track segment finding and track segment association algorithms already use cross-superlayer geometric continuity, hit associations, and track-fitting information. If upstream denoising also introduced such global decisions, it would amount to performing local or even global pattern recognition or fitting before formal track finding. This would not only significantly increase the complexity and computational cost of denoising, but the resulting intermediate information would generally be difficult for subsequent algorithms to reuse directly. Instead, it would consume time and computational resources needed for subsequent reconstruction. Therefore, the functional boundary of upstream denoising should remain distinct from that of subsequent reconstruction: it should remove clearly invalid hits at the earliest stage, rather than perform track recognition in advance.

This boundary is particularly important for the STCF HLT during the R\&D phase. Under the current design condition of a maximum event rate of approximately $400~\mathrm{kHz}$ and with limited GPU resources, the time available for processing one event on one GPU is on the order of $100~\mu\mathrm{s}$. This time budget must cover not only denoising, but also leave sufficient margin for subsequent clustering, two-dimensional track finding, and track fitting. Therefore, the upstream denoising algorithm should keep its operations local, maintain a stable processing flow, and be suitable for execution in parallel on GPUs. Its processing time should not exhibit significant fluctuations with changes in the number of input hits.

Based on the considerations above, this paper defines upstream MDC denoising as a local and conservative background-hit filtering task. Its goal is not to remove every hit labeled as background in a data sample, but to prioritize the removal of isolated points, small irrelevant clusters, and hits that clearly lack continuous local support or are inconsistent with the structure of a local track segment. Hits with track-segment-like shapes, hits that may overlap with genuine tracks, or hits whose physical origin can be determined only through subsequent global reconstruction should be retained whenever possible. In particular, some large hit clusters produced by background processes may not be physically associated with the target process, but directly removing them at the upstream stage could still cause genuine track information to be lost. Such clusters are therefore better handled using global information in subsequent hit association or cross-superlayer reconstruction stages. Based on this positioning, the upstream MDC denoising algorithm should satisfy the following requirements: it should execute efficiently in parallel on GPUs and maintain a low processing time per event; it should retain genuine hits at a ratio as close to $100\%$ as possible for high-momentum tracks, locally complete genuine structures, and reduced detector efficiency; it should primarily remove isolated points, small irrelevant clusters, and background hits lacking continuous support; and it should adopt a conservative retention strategy for hits whose local morphology resembles that of genuine tracks but cannot be reliably classified at the upstream stage. Only under these constraints is upstream denoising suitable for placement at the beginning of the HLT workflow, where it can provide subsequent track finding and reconstruction with input of lower occupancy and higher purity.

\section{Adaptive Matrix-Based Denoising Algorithm Design}

\subsection{Matrix-Based Sliding-Window Model}

The MDC sense wires are arranged in an approximately circular geometry around the coordinate origin and comprise 48 layers of approximately square drift cells. These layers are structurally divided into eight superlayers, with six consecutive layers in each superlayer and the same number of cells in each layer of a given superlayer. The eight circular superlayers are unfolded one by one and represented as eight independent two-dimensional matrices. Each MDC superlayer is represented by a two-dimensional binary matrix with a fixed row number of six, corresponding to the six detector layers in the superlayer. The number of columns is determined by the number of sense wires in the superlayer and ranges from 128 to 352. A matrix element of one indicates that the corresponding sense wire has a valid hit in the current event, whereas zero indicates no hit. The row and column coordinates uniquely identify each sense wire, and relative positions in the matrix are therefore equivalent to the physical relative positions of the sense wires in the MDC.

\begin{figure}[!htb]
	\centering
	\includegraphics[width=0.95\columnwidth]{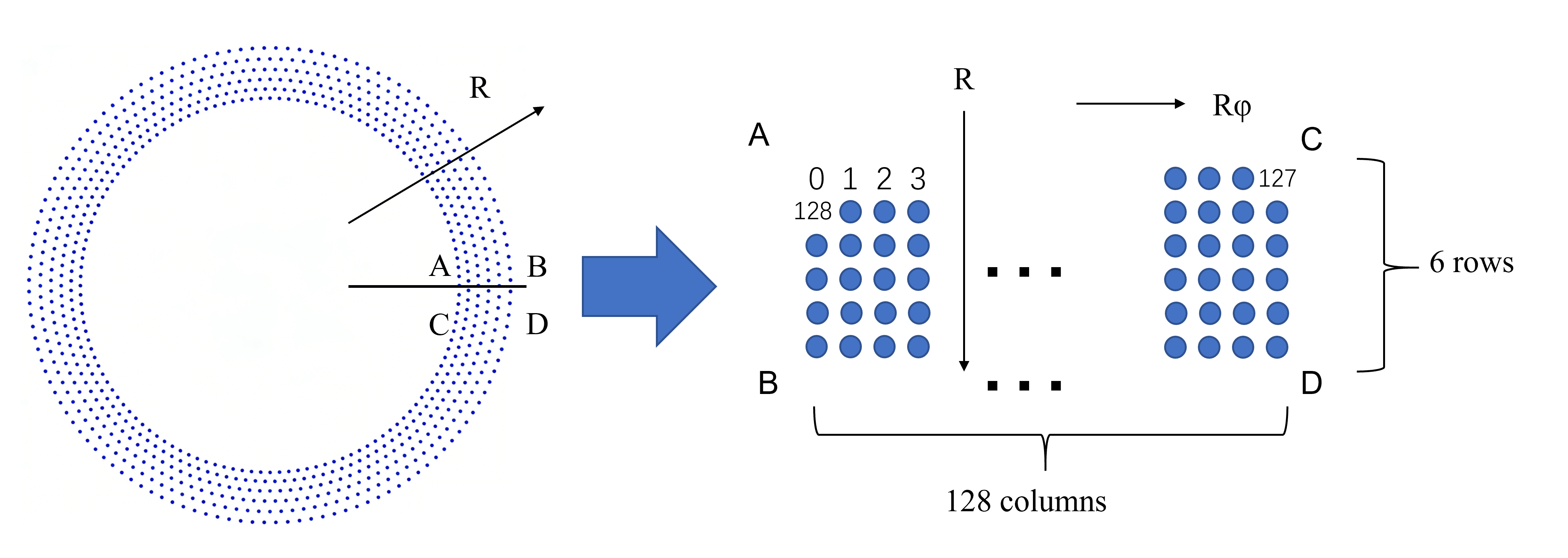}
	\caption{Mapping of all sense wires in the first superlayer to a two-dimensional matrix. The detector-wire IDs in the MDC start from zero and increase counterclockwise. The IDs correspond to the row-major ordering of the elements in the two-dimensional matrix.}
	\label{fig:superlayer_matrix_mapping}
\end{figure}

Because each superlayer is originally circular, unfolding it into a planar two-dimensional matrix may introduce an artificial boundary. If a particle trajectory crosses the positive half of the $x$ axis, close to the unfolding seam, the corresponding track segment may be split between the first few and last few columns of the matrix. To avoid this effect, the first few columns are appended after the last column during data construction, thereby preserving the continuity of trajectories crossing the boundary. When a particle traverses a superlayer, it leaves a corresponding track segment. In the matrix of each superlayer, an element is set to one if the corresponding sense wire is hit and to zero otherwise.

\begin{figure}[!htb]
	\centering
	\includegraphics[width=0.95\columnwidth]{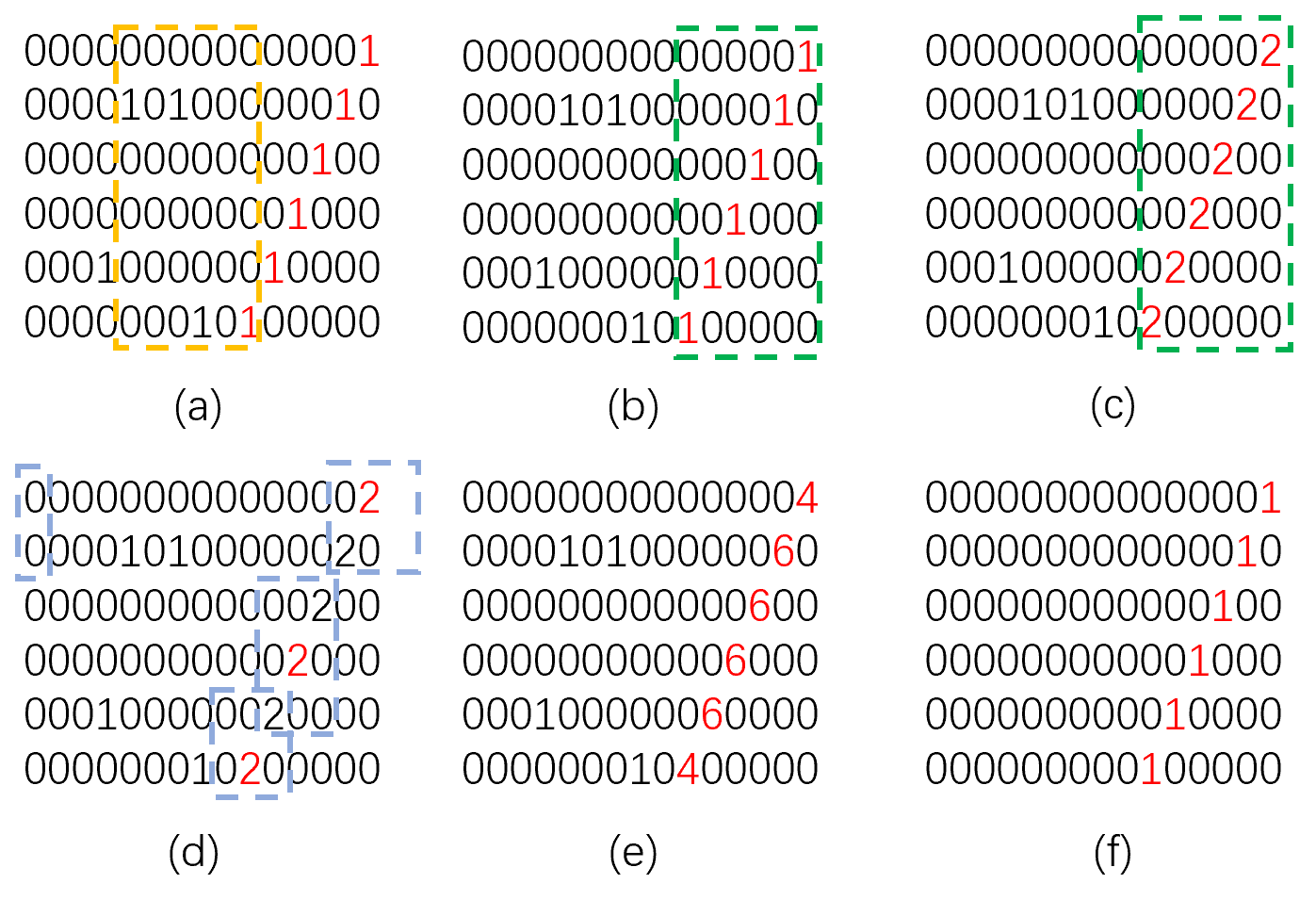}
	\caption{Illustration of the matrix denoising process. In the matrices, red nonzero elements represent hits produced by the target particle, black nonzero elements represent background hits, and zero elements indicate detector wires without hits. The sliding-window size is assumed to be $6\times8$, with $\texttt{threshold}=6$ and $\texttt{limit}=4$. The first step is shown in panels~(a)--(c), the second step in panels~(d)--(e), and the third step in panel~(f). Through these three steps, discrete point clusters are removed while the particle track segment is retained.}
	\label{fig:matrix_denoising_process}
\end{figure}

In form, the algorithm resembles a local feature-extraction procedure. It is inspired by the concepts of a local receptive field and spatial-feature enhancement, but avoids neural-network training and inference, does not require training, introduces no floating-point matrix multiplication, and does not depend on global track fitting. The implementation consists of three simple and well-defined steps, each of which is naturally suited to highly parallel execution. The overall procedure is illustrated in Fig.~\ref{fig:matrix_denoising_process}.

Here and below, \texttt{threshold} and \texttt{limit} denote the names of program parameters, rather than uses of ``threshold'' and ``limit'' as ordinary descriptive terms in the prose.

The first step is sliding-window accumulation. For each superlayer matrix, a $6\times 8$ sliding window is constructed and moved across the entire superlayer along the column direction. For each window, the number of nonzero elements is counted. If the number of hits in the window reaches the specified threshold, the local region covered by the window is considered to possess a certain degree of spatial correlation, and every nonzero hit in the window receives one support increment. If the number of hits is below the threshold, the window provides no support to its hits. Because the original MDC superlayer is circular and periodic in the azimuthal direction, the sliding window obeys a periodic boundary condition along the column direction. When the window extends beyond the last matrix column, the out-of-range part is mapped back to the first columns. Through this step, hits in locally dense regions are covered repeatedly by multiple active windows and therefore obtain higher support values, whereas isolated hits or a small number of background points usually receive fewer increments because they rarely satisfy the window condition.

The second step is local-neighborhood enhancement. For the matrix obtained after the first step, a $3\times 3$ neighborhood is defined around each nonzero element, and the sum of the values in this neighborhood is taken as the new support value of the center point. The row direction follows the actual matrix boundary, whereas the column direction retains the periodic boundary condition of the MDC. This step further exploits the near-neighbor spatial relationships among hits, allowing adjacent hits belonging to the same local track segment to enhance one another. If several supported neighboring points surround a hit, its final score increases further; if a hit lacks other supported points in its neighborhood, its score remains relatively low after enhancement.

The third step is threshold filtering. The enhanced matrix is traversed element by element. If an element has a value greater than or equal to the final threshold $L$, it is set to one and retained; otherwise, it is set to zero. The matrix is thus restored to a binary hit matrix. Retained elements correspond to MDC hits considered to have sufficient local spatial support, whereas zeroed elements correspond to hits with insufficient local support and are more likely to be isolated background hits or members of small background clusters. The matrix-based denoising algorithm requires neither complete track parameters nor global fitting. It relies only on the positional relationships of nonzero elements within a superlayer matrix, making its computational procedure simple and regular and therefore suitable for parallel implementation.

\begin{figure}[!htb]
	\centering
	\includegraphics[width=0.95\columnwidth]{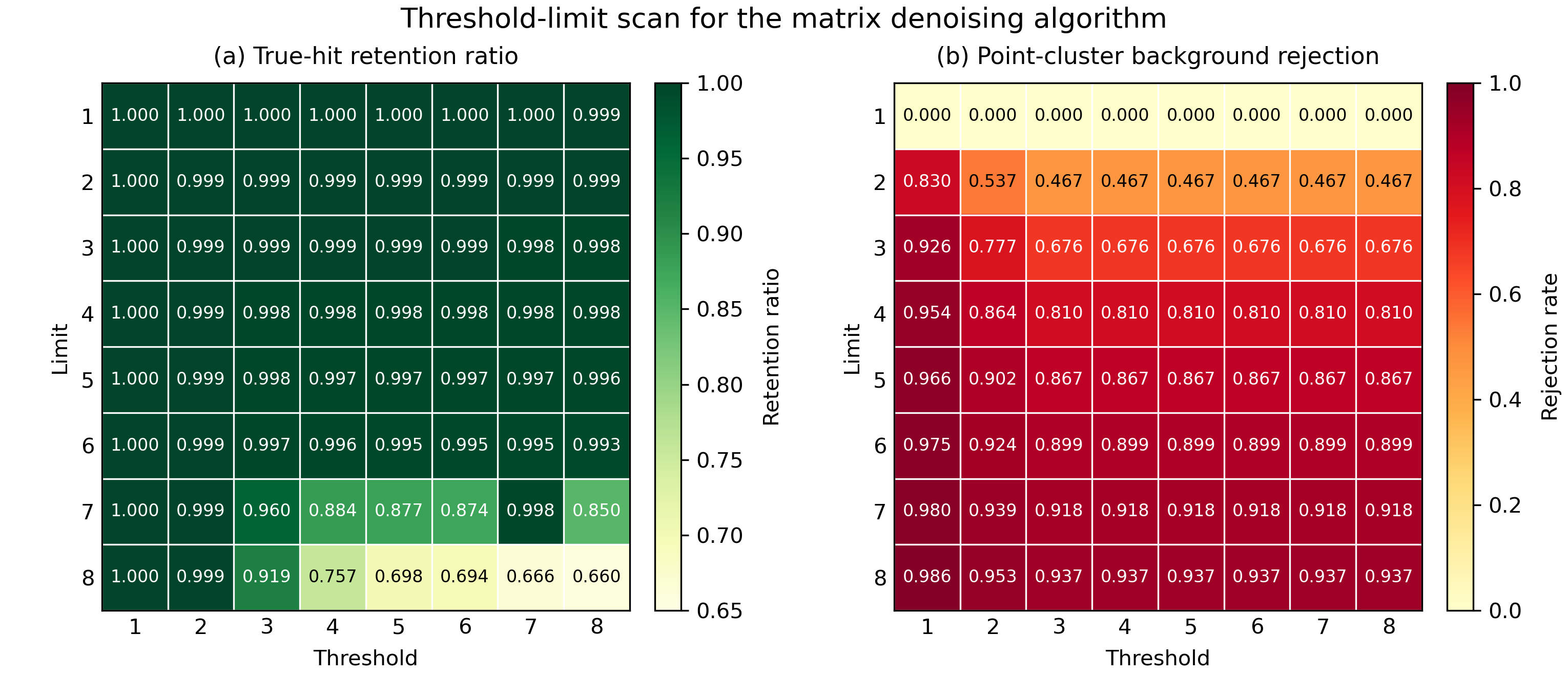}
\caption{Heatmaps of the true-hit retention ratio and point-cluster background rejection rate for the \(e^+e^- \to \pi^+\pi^- J/\psi,\ J/\psi \to e^+e^-\) channel under different combinations of \texttt{threshold} and \texttt{limit}. The result for each cell is obtained from $10\,\mathrm{k}$ events. Different combinations of \texttt{threshold} and \texttt{limit} yield different performance metrics, and an appropriate combination can be selected according to the user's requirements. The horizontal axis represents \texttt{threshold}, and the vertical axis represents \texttt{limit}.}
	\label{fig:threshold_limit_heatmap}
\end{figure}

The algorithm has three main parameters: the sliding-window size, the window-statistics threshold \texttt{threshold}, and the final binarization threshold \texttt{limit}. The sliding-window size is generally determined by the morphology of a particle track segment in the two-dimensional superlayer matrix. Because \texttt{threshold} and \texttt{limit} are discrete integer parameters, their combinations can be scanned using simulation samples or calibration samples extracted during operation. A suitable parameter combination can then be selected using the true-hit retention ratio, background rejection rate, and algorithm processing time as performance metrics. Fig.~\ref{fig:threshold_limit_heatmap} shows the true-hit retention ratio and point-cluster background suppression rate obtained by scanning different combinations of \texttt{threshold} and \texttt{limit} for the \(e^+e^- \to \pi^+\pi^- J/\psi,\ J/\psi \to e^+e^-\) channel with one nominal background level. Each parameter point is evaluated using $10\,\mathrm{k}$ events.

\subsection{Adaptive Threshold Selection}

During actual STCF operation, the best parameter combination determined from simulation could be selected in advance according to the current background level and detector status, allowing the algorithm to maintain good denoising performance under different operating conditions. However, the adaptive capability can be further enhanced by automatically adjusting the \texttt{threshold} and \texttt{limit} combination according to the background level of each input event, so that the algorithm can achieve the best possible denoising performance over a range of background conditions.

The central idea of the adaptive threshold-selection mechanism is to preserve the main workflow of the fixed-threshold matrix-based denoising algorithm---sliding-window accumulation, local-neighborhood enhancement, and threshold filtering---while introducing event-level adaptation only in the parameter-selection stage. First, the global occupancy $p$ of the current event is used to select the sliding-window threshold \texttt{threshold}. During sliding-window accumulation, the number of active windows is counted simultaneously and the window activation ratio $n$ is calculated. The final filtering threshold \texttt{limit} is no longer fixed, but is determined by an offline-calibrated function $f(p,n)$. During online operation, the algorithm only needs to calculate $p$ and $n$ for the current event and then obtain \texttt{limit} from an offline-calibrated polynomial threshold-mapping model. This avoids the additional cost of scanning candidate parameters online while improving the adaptability to different background conditions.

The global occupancy of the current event is defined as
\begin{equation}
p=\frac{N_{\text{hit}}}{N_{\text{wire}}},
\label{eq:event_occupancy}
\end{equation}
where $N_{\text{hit}}$ is the number of hit sense wires in the current event, including background hits, and $N_{\text{wire}}$ is the total number of sense wires in the MDC. The quantity $p$ characterizes the overall hit density of the event. Under normal conditions with stable channels, a larger $p$ generally indicates a higher overall occupancy and a greater likelihood of abundant background hits; the activation condition for a sliding window should therefore be relatively strict. A smaller $p$ indicates a sparser input, for which the window threshold can be reduced appropriately to avoid prematurely weakening genuine but locally incomplete track segments.

The threshold used in the first sliding-window step can be written as
\begin{equation}
T=\mathcal{P}(p),
\label{eq:window_threshold}
\end{equation}
where $T$ is the threshold used during the first sliding-window accumulation, corresponding to the program parameter \texttt{threshold}, and $\mathcal{P}(p)$ is the mapping from event occupancy to the window threshold. In practice, $\mathcal{P}$ can be represented as a piecewise-constant function:
\begin{equation}
\mathcal{P}(p)=\tau_k,\qquad p\in I_k,\qquad k=1,2,\ldots,K,
\label{eq:piecewise_threshold}
\end{equation}
where $\{I_k\}_{k=1}^{K}$ are intervals obtained by partitioning the occupancy distribution of offline samples, and $\tau_k$ is the sliding-window threshold used in the corresponding interval. For a given event, the algorithm determines \texttt{threshold} from the interval containing its occupancy $p$ and does not need to search for the threshold online.

After $T$ has been determined, the algorithm performs sliding-window accumulation according to the original procedure. The number of active windows is counted simultaneously, with nearly no additional overhead. The window activation ratio is defined as
\begin{equation}
n=\frac{N_{\text{active win}}}{N_{\text{all win}}},
\label{eq:window_activation_ratio}
\end{equation}
where $N_{\text{active win}}$ is the number of active sliding windows in the current event, namely, the number of windows whose number of nonzero elements reaches the threshold $T$, and $N_{\text{all win}}$ is the total number of sliding windows, which is fixed for the algorithm configuration used in this paper. Compared with $p$, $n$ is more directly related to the internal state of the algorithm. Two events may have similar overall occupancy, but if the hits in one event are dispersed and those in the other are concentrated, their window activation ratios can differ substantially. Thus, $p$ describes the overall input intensity, whereas $n$ describes the local aggregation produced by the window scan; both are used as inputs to the final threshold.

Let the final filtering threshold, corresponding to the program parameter \texttt{limit}, be denoted by $L$. In the adaptive algorithm, $L$ is jointly determined by the global occupancy $p$ and the window activation ratio $n$. The continuous prediction is written as
\begin{equation}
L_0=f(p,n).
\label{eq:continuous_limit}
\end{equation}
To keep the final threshold within a reasonable integer range, the continuous prediction is rounded and range-limited in actual use:
\begin{equation}
L=\operatorname{clip}\left(\left\lfloor L_0+\frac{1}{2}\right\rfloor,L_{\min},L_{\max}\right),
\label{eq:final_limit}
\end{equation}
where $\lfloor L_0+1/2\rfloor$ denotes rounding the continuous prediction to the nearest integer, and $\operatorname{clip}(x,L_{\min},L_{\max})$ restricts $x$ to the interval $[L_{\min},L_{\max}]$.

In the adaptive algorithm implemented in this work, the continuous function $f(p,n)$ is represented by a fourth-degree polynomial. To reduce the effect of different feature scales on the model output, $p$ and $n$ are first standardized using the training samples:
\begin{equation}
\hat p=\frac{p-\mu_p}{\sigma_p},\qquad
\hat n=\frac{n-\mu_n}{\sigma_n},
\label{eq:standardized_features}
\end{equation}
where $(\mu_p,\sigma_p)$ are the mean and standard deviation of $p$ in the training samples, and $(\mu_n,\sigma_n)$ are the corresponding quantities for $n$. The function $f(p,n)$ is then written as a fourth-degree polynomial in the standardized features:
\begin{equation}
f(p,n)=\sum_{i+j\le 4}a_{ij}\hat p^{\,i}\hat n^{\,j},\qquad i,j\ge 0,
\label{eq:fourth_order_model}
\end{equation}
where $a_{ij}$ are polynomial coefficients obtained by offline calibration. The expression contains all terms with total degree no greater than four, including the constant, first-, second-, third-, and fourth-degree terms, as well as cross terms between $p$ and $n$.

The process described above is the forward calculation performed during online operation. The coefficients of the fourth-degree polynomial are not manually set online; they are determined through offline calibration. During offline calibration, a set of candidate \texttt{limit} values is scanned for the calibration samples, and the true-hit retention ratio and discrete-background rejection rate are measured for each threshold. For a given event, subject to a prescribed lower bound on the true-hit retention ratio, the candidate threshold with the best discrete-background rejection performance is selected and denoted by $L^\ast$. Thus, each calibration event provides an input--output pair mapping its $(p,n)$ to the reference optimal threshold $L^\ast$.

After the calibration samples have been obtained, $p$ and $n$ are standardized for all samples and polynomial features with total degree no greater than four are constructed. The coefficients $a_{ij}$ are then determined by least squares with a weak ridge-regularization term, using $L^\ast$ as the fitting target. After fitting, all standardization parameters and polynomial coefficients are fixed in the online program. The online stage no longer scans candidate \texttt{limit} values; it directly calculates $f(p,n)$ from the current event's $p$ and $n$, and obtains the final threshold $L$ through rounding and range limiting.

The offline calibration samples combine the five physics channels considered in this work, rather than establishing an independent parameter set for each channel. Consequently, the standardization parameters, polynomial coefficients, and threshold-selection rules are shared by all channels. During online operation, the occupancy $p$ and the window activation ratio $n$ are calculated independently for each event, and the corresponding threshold is determined separately for that event. If multiple events are processed in a GPU batch, the batch is used only as a computational organization unit; the occupancy and threshold of one event are not obtained by averaging or combining the hits of other events. The validation results obtained with the same parameter set across the five physics channels indicate that the calibrated algorithm has a certain degree of adaptability and robustness to differences in event topology and hit multiplicity. The corresponding comparative results are presented in the next section.

Because model training and parameter calibration are completed offline, they introduce no online search overhead. During actual online operation, the adaptive algorithm only needs to perform one additional low-order polynomial forward calculation to obtain the \texttt{limit} for the current event. The adaptive threshold-selection mechanism therefore enhances the adaptability of the denoising algorithm while maintaining low additional computational cost. The choice of the functional form of $f(p,n)$ is discussed further in the next section.


\section{Algorithm Performance}

The performance of the proposed algorithm is evaluated from two aspects: physics performance and deployment performance on the HLT prototype. The physics performance includes the signal retention ratio, output hit fraction, and point-cluster background rejection rate. The deployment performance includes GPU processing time, batch-processing throughput, end-to-end latency, and resource utilization.

\subsection{Physics Metrics and Evaluation}

The Offline Software of the Super Tau-Charm Facility (OSCAR)~\cite{OSCAR1,OSCAR2} is mainly used for offline data processing, including detector simulation, digitization, reconstruction, and physics analysis. We use OSCAR to generate five physics-process samples containing charged final states. The covered energy points include the $J/\psi$ peak at $3.097~\mathrm{GeV}$, as well as $4.26~\mathrm{GeV}$ and $4.682~\mathrm{GeV}$, which are relevant to other physics processes. The momentum range of the final-state particles is approximately $[0.1,2]~\mathrm{GeV}/c$, covering the medium- and high-momentum regions. Their spatial distribution covers the polar-angle range of $[20^\circ,160^\circ]$, corresponding to the detector acceptance. Using OSCAR, we generated events with different background levels. The typical backgrounds primarily consist of two categories: beam-related backgrounds and luminosity-related backgrounds. Beam-related background mainly includes the Touschek effect and beam-gas interactions (Coulomb scattering and bremsstrahlung radiation); luminosity-related background mainly includes radiative Bhabha scattering (RBB) and two-photon processes.The two components contribute with a typical ratio of approximately $6:1$\cite{OSCAR3}. Table~\ref{tab:hit_occupancy} shows the ratio between the signals produced in the MDC by high-energy physics particles and the background signals produced in the MDC in the datasets used in this work. In the following, $N\mathrm{Bkg}$, with $N=1$, $2$, $3$, or $5$, denotes the corresponding background level.

\begin{table*}[t]
	\centering
	\caption{MDC hit occupancy statistics for five physics channels under different background levels. The real-hit ratio is defined as the ratio of hits from true particle tracks to all hits. The hit occupancy ratio is defined as the ratio of the number of hit sense wires in each event to the total number of sense wires. The values are averaged over \(10^4\) events.}
	\label{tab:hit_occupancy}
	\resizebox{\textwidth}{!}{
		\begin{tabular}{lcccccccc}
			\toprule
			\multirow{2}{*}{Physics channel}
			& \multicolumn{2}{c}{\(1{\rm Bkg}\)}
			& \multicolumn{2}{c}{\(2{\rm Bkg}\)}
			& \multicolumn{2}{c}{\(3{\rm Bkg}\)}
			& \multicolumn{2}{c}{\(5{\rm Bkg}\)} \\
			\cmidrule(lr){2-3}
			\cmidrule(lr){4-5}
			\cmidrule(lr){6-7}
			\cmidrule(lr){8-9}
			& Real-hit ratio & Hit occupancy ratio
			& Real-hit ratio & Hit occupancy ratio
			& Real-hit ratio & Hit occupancy ratio
			& Real-hit ratio & Hit occupancy ratio \\
			\midrule
			\makecell[l]{\(e^+e^- \to \pi^+\pi^- J/\psi\),\\\(J/\psi \to e^+e^-\)}
			& 0.270 & 0.066
			& 0.161 & 0.111
			& 0.117 & 0.153
			& 0.078 & 0.229 \\
			
			\makecell[l]{\(e^+e^- \to \pi^+\pi^- J/\psi\),\\\(J/\psi \to \mu^+\mu^-\)}
			& 0.264 & 0.067
			& 0.159 & 0.111
			& 0.115 & 0.153
			& 0.077 & 0.228 \\
			
			\makecell[l]{\(e^+e^- \to K^+K^- J/\psi\),\\\(J/\psi \to l^+l^-\)}
			& 0.303 & 0.070
			& 0.184 & 0.115
			& 0.136 & 0.156
			& 0.091 & 0.231 \\
			
			\(e^+e^- \to \tau^+\tau^-\)
			& 0.156 & 0.058
			& 0.087 & 0.103
			& 0.062 & 0.146
			& 0.041 & 0.222 \\
			
			\(J/\psi \to \Lambda\bar{\Lambda}\)
			& 0.344 & 0.074
			& 0.214 & 0.118
			& 0.158 & 0.160
			& 0.108 & 0.234 \\
			\bottomrule
		\end{tabular}
	}
\end{table*}

The physics performance of the denoising algorithm is evaluated using the signal retention ratio, the output hit fraction, and the point-cluster background rejection rate.

The signal retention ratio is defined as the ratio of the number of true-track hits retained by the algorithm to the number of input true-track hits:
\begin{equation}
	\epsilon_{\rm sig}
	=
	\frac{N_{\rm true}^{\rm out}}
	{N_{\rm true}^{\rm in}} .
	\label{eq:signal_retention_ratio}
\end{equation}

The output hit fraction is defined as the ratio of the total number of output hits to the total number of input hits:
\begin{equation}
	f_{\rm out}
	=
	\frac{N_{\rm hit}^{\rm out}}
	{N_{\rm hit}^{\rm in}} .
	\label{eq:output_hit_fraction}
\end{equation}

The point-cluster background rejection rate is defined as the fraction of input point-cluster background hits that are removed by the algorithm:
\begin{equation}
	r_{\rm pc}
	=
	1-
	\frac{N_{\rm pc}^{\rm out}}
	{N_{\rm pc}^{\rm in}} .
	\label{eq:point_cluster_rejection}
\end{equation}

Here, $N_{\rm true}^{\rm in}$ and $N_{\rm true}^{\rm out}$ denote the numbers of true-track hits before and after denoising, respectively. $N_{\rm hit}^{\rm in}$ and $N_{\rm hit}^{\rm out}$ denote the total numbers of input and output hits, respectively. $N_{\rm pc}^{\rm in}$ and $N_{\rm pc}^{\rm out}$ denote the numbers of hits belonging to point-cluster backgrounds before and after denoising, respectively. In this work, point-cluster backgrounds are defined as discrete clusters containing one to three adjacent background hits.

In this work, discrete point clusters consisting of one to three adjacent background hits are used as the statistical objects for evaluating the point-cluster background rejection rate, and the number of hits contained in these clusters is counted. When the detector efficiency $\eta$ is considered, finite detection efficiency is simulated by randomly removing a fraction $1-\eta$ of the genuine-track hits and background hits. The input hit counts used as denominators in the above metrics are the hits that actually enter the algorithm after the detector-efficiency procedure. Therefore, the signal retention ratio, output hit fraction, and point-cluster background rejection rate all lie between zero and one. In all subsequent experiments in this paper, no time window is applied; the event start-time range is $0$--$2000~\mathrm{ns}$. 
Since any data that need to be stored must pass through the HLT, data rejected by the HLT will not be retained. To ensure the accuracy of offline reconstruction, the signal retention ratio of the denoising algorithm should be as close to $100\%$ as possible. The point-cluster background rejection rate should be maximized while maintaining a high signal retention ratio. For the reasons described above, our algorithm does not specifically pursue an extremely low output hit fraction. Instead, this quantity is provided for reference, allowing the fraction of the input signals remaining after point-cluster removal to be directly observed.

For the most important metrics, namely the signal retention ratio and point-cluster background rejection rate, we generated five physics channels using OSCAR, with $10\,\mathrm{k}$ events for each channel. Table~\ref{tab:denoising_physics_performance} presents the average signal retention ratio, output hit fraction, and point-cluster background rejection rate obtained after processing $10\,\mathrm{k}$ events with different background levels. The point-cluster background rejection rate exceeds $92\%$ from $1$ to $3\mathrm{Bkg}$ and decreases to $79.49\%$--$80.70\%$ at $5\mathrm{Bkg}$. The signal retention ratio remains above $98.48\%$ for all physics channels and background levels. As the background level increases, the output hit fraction rises from $74.26\%$--$80.44\%$ at $1\mathrm{Bkg}$ to $85.56\%$--$87.12\%$ at $5\mathrm{Bkg}$, indicating that a larger fraction of complex background hits is conservatively retained under higher-occupancy conditions.

\begin{table*}[t]
	\centering
	\caption{Physics performance of the adaptive matrix-based denoising algorithm for five physics channels under different background levels. The point-cluster background rejection rate, signal retention ratio, and output hit fraction are listed for \(1{\rm Bkg}\), \(2{\rm Bkg}\), \(3{\rm Bkg}\), and \(5{\rm Bkg}\). The channel order is consistent with Table~\ref{tab:hit_occupancy}.}
	\label{tab:denoising_physics_performance}
	\begin{tabular*}{\textwidth}{@{\extracolsep{\fill}}llcccc}
		\toprule
		Physics channel
		& Metric
		& \(1{\rm Bkg}\)
		& \(2{\rm Bkg}\)
		& \(3{\rm Bkg}\)
		& \(5{\rm Bkg}\) \\
		\midrule
		
		\multirow{3}{*}{\makecell[l]{\(e^+e^- \to \pi^+\pi^- J/\psi\),\\\(J/\psi \to e^+e^-\)}}
		& Point-cluster background rejection rate
		& 96.25\% & 95.19\% & 92.82\% & 79.66\% \\
		& Signal retention ratio
		& 99.29\% & 99.31\% & 99.44\% & 99.14\% \\
		& Output hit fraction
		& 78.53\% & 79.00\% & 81.13\% & 86.64\% \\
		\midrule
		
		\multirow{3}{*}{\makecell[l]{\(e^+e^- \to \pi^+\pi^- J/\psi\),\\\(J/\psi \to \mu^+\mu^-\)}}
		& Point-cluster background rejection rate
		& 96.28\% & 95.08\% & 92.98\% & 79.77\% \\
		& Signal retention ratio
		& 99.31\% & 99.42\% & 99.45\% & 99.14\% \\
		& Output hit fraction
		& 78.65\% & 78.75\% & 81.13\% & 86.58\% \\
		\midrule
		
		\multirow{3}{*}{\makecell[l]{\(e^+e^- \to K^+K^- J/\psi\),\\\(J/\psi \to l^+l^-\)}}
		& Point-cluster background rejection rate
		& 96.13\% & 94.69\% & 92.67\% & 79.49\% \\
		& Signal retention ratio
		& 99.19\% & 99.21\% & 99.30\% & 99.24\% \\
		& Output hit fraction
		& 79.74\% & 79.66\% & 81.58\% & 86.91\% \\
		\midrule
		
		\multirow{3}{*}{\(e^+e^- \to \tau^+\tau^-\)}
		& Point-cluster background rejection rate
		& 96.67\% & 94.99\% & 93.23\% & 80.70\% \\
		& Signal retention ratio
		& 99.33\% & 99.43\% & 99.53\% & 99.26\% \\
		& Output hit fraction
		& 74.26\% & 76.49\% & 79.23\% & 85.56\% \\
		\midrule
		
		\multirow{3}{*}{\(J/\psi \to \Lambda\bar{\Lambda}\)}
		& Point-cluster background rejection rate
		& 96.17\% & 94.76\% & 92.16\% & 79.60\% \\
		& Signal retention ratio
		& 98.56\% & 98.48\% & 98.81\% & 99.28\% \\
		& Output hit fraction
		& 80.44\% & 80.25\% & 82.16\% & 87.12\% \\
		
		\bottomrule
	\end{tabular*}
\end{table*}
To examine the influence of the denoising algorithm on tracks with different transverse momenta, we used OSCAR to generate single-particle $\mu^+$ samples. These samples were generated with a particle gun by firing one $\mu^+$ with a specified transverse momentum into the detector. The azimuthal angle was uniformly distributed over $0^\circ$--$360^\circ$, while the polar angle was fixed to $30^\circ$, $60^\circ$, or $90^\circ$. The transverse-momentum range was $50$--$1500~\mathrm{MeV}/c$. One nominal background level was mixed into these events. For tracks with transverse momenta in this range, the algorithm retains the tracks effectively, without significant fluctuations in the signal retention ratio caused by changes in transverse momentum.

\begin{figure}[!htb]
	\centering
	\includegraphics[width=0.95\columnwidth]{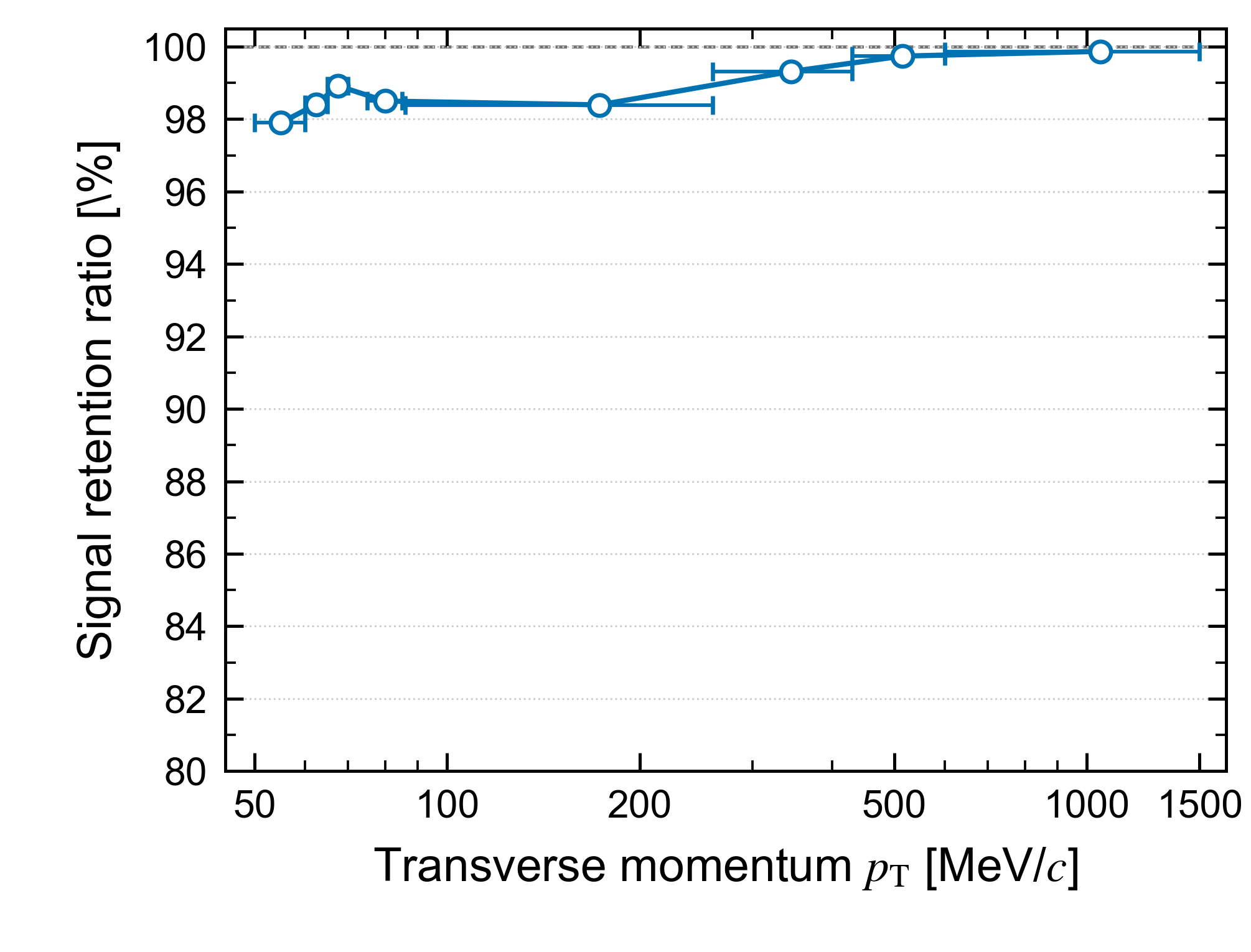}
\caption{Retention capability of the denoising algorithm for particle tracks with different transverse momenta. The horizontal axis represents the transverse momentum, and the vertical axis represents the signal retention ratio. Single-particle $\mu^+$ events are used, with transverse momenta ranging from $50$ to $1500~\mathrm{MeV}/c$ and one nominal background level mixed into the samples.}
	\label{fig:signal_retention_vs_pt}
\end{figure}

Detector efficiency is another important indicator of the robustness of a denoising algorithm. As discussed in Section~2, finite detector efficiency may cause local breaks in track segments represented in the two-dimensional superlayer matrices, causing them to degenerate into several clusters consisting of only a few isolated hits. Therefore, the effect of detector efficiency must be fully considered when designing an MDC denoising algorithm. A simple noise-removal strategy may inevitably remove some genuine charged-particle hits together with the noise, thereby reducing the signal retention ratio.

To evaluate the robustness of the proposed algorithm, we designed a set of comparative experiments. All denoising algorithms used in these experiments rely only on information within a single superlayer and do not use cross-superlayer information.

%
%

\noindent\textbf{Single-hit denoising.}
This is the most basic filtering method based on absolute spatial isolation.
If a detector hit has no other hit within its $3\times3$ neighborhood in the
two-dimensional matrix, it is classified as random noise and removed.

\medskip

\noindent\textbf{Cluster denoising.}
This method is based on a local connected-component search. Spatially adjacent
hits are grouped into clusters, and clusters containing no more than three hits
are classified as noise and removed. Specifically, the $3\times3$ region around
each hit is searched. If another hit is found, the hits are considered connected.
A connected cluster containing no more than three hits is filtered out. This
method partially compensates for the limitations of single-hit denoising by
removing clusters formed when random noise hits happen to be connected.

\medskip

\noindent\textbf{Morphological denoising.}
This method extends single-hit denoising using a structuring element defined
according to the geometry of the superlayer. Specifically, for a rectangular
sliding window, if the outer ring contains no hits, all elements in the inner
region are set to zero.

\begin{figure}[!htb]
	\centering
	\includegraphics[width=0.95\columnwidth]{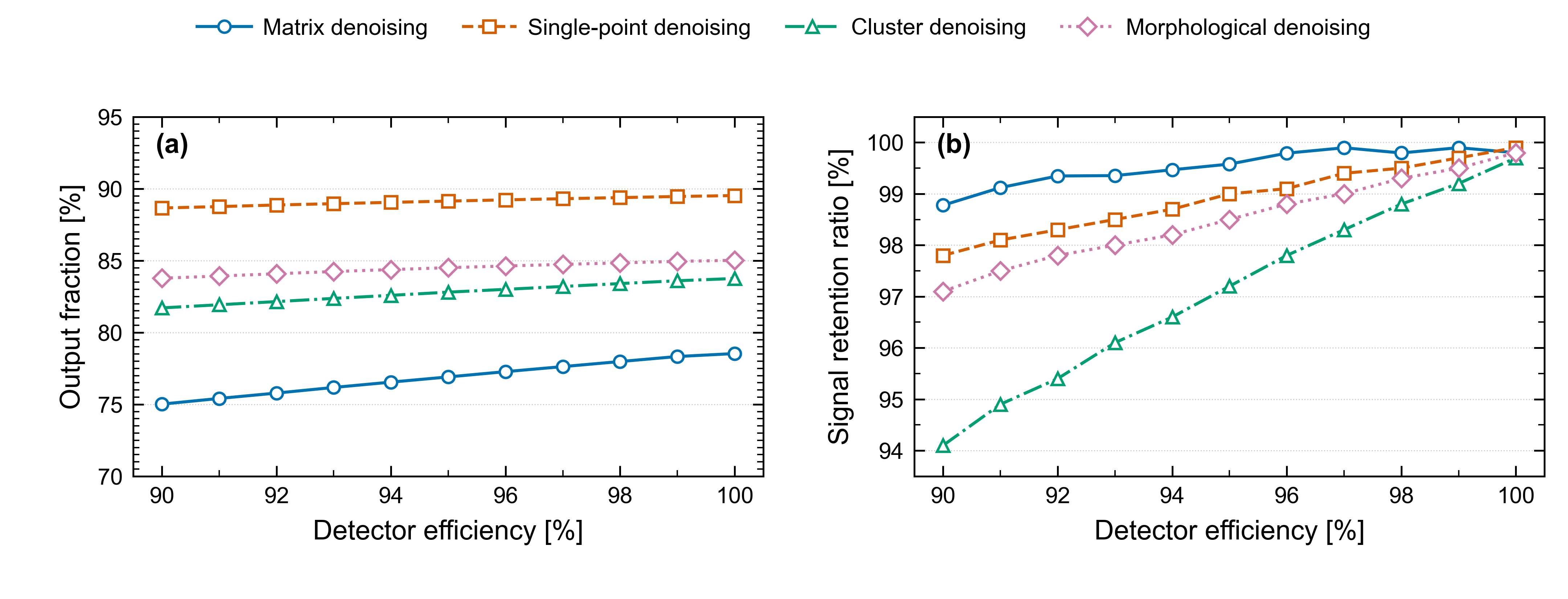}
	\caption{Performance comparison of different denoising algorithms under the $1\mathrm{Bkg}$ \(e^+e^- \to \pi^+\pi^- J/\psi,\ J/\psi \to e^+e^-\) channel. The left panel shows the output hit fraction as a function of the detector efficiency, while the right panel shows the signal retention ratio as a function of the detector efficiency. In the detector-efficiency range of $90\%$--$100\%$, the adaptive matrix-based denoising algorithm achieves a lower output hit fraction than the other algorithms, while maintaining a signal retention ratio above $98\%$.}
	\label{fig:efficiency_algorithm_comparison}
\end{figure}

Because the denoising algorithm must accommodate a wide range of possible particle-track morphologies, we generated events with one nominal background level and transverse momenta in the range of $50$--$1500~\mathrm{MeV}/c$ using OSCAR. The events were processed separately by the four algorithms, and the resulting signal retention ratios and point-cluster background rejection rates were calculated. The results are shown in Fig.~\ref{fig:efficiency_algorithm_comparison}.

To verify the effectiveness of the polynomial mapping function proposed in this work, three common regression models were additionally introduced for comparison. All four methods use the event features $p$ and $n$ as inputs and the optimal \texttt{limit} as the regression target. The predicted \texttt{limit} is then applied to the same subsequent denoising procedure for evaluation.

The first method is the polynomial mapping function proposed in this work. It constructs a continuous mapping between $p$, $n$, their combinations at different orders, and the optimal \texttt{limit}. The polynomial describes the main trend of \texttt{limit} as a function of event occupancy and local aggregation, while maintaining a simple expression with a small number of parameters. Its evaluation requires only the substitution of the input features into the polynomial coefficients.

The second method is random-forest regression. A random forest constructs multiple regression decision trees and combines their predictions to describe the nonlinear relationship between $(p,n)$ and \texttt{limit}.

The third method is radial-basis-function (RBF) regression. This method uses radial basis functions to describe the distances between test samples and the centers of training samples, and uses kernel functions to construct a nonlinear mapping in the input-feature space. RBF regression has good local fitting capability and can model different occupancy-feature regions nonlinearly.

The fourth method is multilayer perceptron (MLP) neural-network regression. An MLP uses nonlinear activation functions in hidden layers to learn the complex relationship between $(p,n)$ and \texttt{limit}, and has a strong general function-approximation capability.

\begin{table}[t]
	\centering
	\caption{Mapping time and denoising performance for the \(e^+e^- \to \pi^+\pi^- J/\psi,\ J/\psi \to e^+e^-\) channel. Results are averaged over \(10\,\mathrm{k}\) events.}
	\label{tab:mapping_method_comparison}
	\setlength{\tabcolsep}{3pt}
	\begin{tabular}{lcccc}
		\toprule
		Metric
		& Polynomial
		& RBF
		& \shortstack{Random\\forest}
		& MLP \\
		\midrule
		\makecell[l]{Mapping time\\per event}
		& $0.194~\mu\mathrm{s}$
		& $0.703~\mu\mathrm{s}$
		& $1.251~\mu\mathrm{s}$
		& $2.133~\mu\mathrm{s}$ \\
		\midrule
		\makecell[l]{Signal retention\\ratio}
		& 0.991
		& 0.988
		& 0.991
		& 0.991 \\
		\midrule
		\makecell[l]{Output hit\\fraction}
		& 0.735
		& 0.736
		& 0.734
		& 0.732 \\
		\midrule
		\makecell[l]{Point-cluster\\ background\\rejection rate}
		& 0.971
		& 0.969
		& 0.973
		& 0.975 \\
		\bottomrule
	\end{tabular}
	
	\vspace{2pt}
	\begin{minipage}{\columnwidth}
		\footnotesize
		The comparison focuses primarily on mapping time and signal retention ratio. The training weights therefore differ slightly from those used for Table~\ref{tab:denoising_physics_performance}, resulting in small differences in the final performance metrics. All methods were evaluated on a single Zen~2 CPU core without hyper-threading, at a fixed frequency of $2.5~\mathrm{GHz}$.
	\end{minipage}
\end{table}

Table~\ref{tab:mapping_method_comparison} presents the processing time and key metrics of the four methods. The random-forest model consists of 32 regression trees. The maximum depth of each tree is set to 10, and each leaf node contains at least 16 samples. During training, training samples are randomly selected using bootstrap sampling, and the two input features are used for node splitting. The RBF model selects 512 radial-basis centers in the standardized two-dimensional input space, with each center corresponding to a Gaussian kernel. The MLP model has two input features corresponding to $p$ and $n$, one hidden layer containing 256 neurons, and the hyperbolic tangent function as the activation function. The output layer contains one neuron for predicting the continuous \texttt{limit} value.

The three supplementary regression models achieve signal retention and noise-suppression performance comparable to those of the polynomial method, indicating that the polynomial fitting function provides sufficient mapping accuracy for the present adaptive denoising task. However, random forests, RBF regression, and MLP regression introduce relatively high additional computational costs and therefore increase the mapping time. The polynomial function used in this work maintains the filtering performance while providing lower computational overhead and better suitability for online deployment.

In two-dimensional MDC reconstruction, track-finding and reconstruction algorithms can be divided into two categories according to their dependence on the input-signal size. The first category is represented by pattern-matching algorithms, whose processing time is mainly determined by the number of candidate tracks or the complexity of local combinations and is relatively insensitive to the total number of input hits. The second category is represented by Hough-transform algorithms, whose computational cost generally increases substantially with the number of input hits. For the latter category, reducing the number of input hits during preprocessing helps reduce the search space and improve processing efficiency.

Because the denoising algorithm is designed primarily for online use in the HLT, its output is used only to filter the data during online processing. Offline reconstruction does not use the reconstruction results produced online; instead, it performs reconstruction independently. Therefore, the benefits and losses introduced by the denoising algorithm for subsequent algorithms should be evaluated using the existing online reconstruction algorithms rather than offline reconstruction algorithms. For this reason, we do not evaluate the denoised structures in the offline system. Instead, we use an already developed online reconstruction algorithm for evaluation.

The Hough transform is a widely used method in image processing and computer vision for detecting geometric shapes, particularly lines and circles, and it has also been widely applied in high-energy physics experiments\cite{ZhouHough2025}. Hough-transform reconstruction requires accumulation and peak finding in Hough space. A circle is determined by locating the maximum in the accumulated distribution. When each hit is mapped into Hough space, a large number of calculations and mapping operations are required, resulting in substantial computational cost. Because the Hough transform is applied to all input hits, reducing the number of input signal points can significantly reduce the time required to search for candidate tracks. In addition, the presence of a small number of isolated points may affect the final circle identified during peak finding.

Accordingly, we performed two sets of experiments:

\begin{enumerate}
	\item the effect of denoising on the processing time of subsequent two-dimensional track finding based on the Hough transform;
	\item the effect of denoising on the track-finding efficiency when the genuine tracks retained by the denoising algorithm are processed using the Hough transform for two-dimensional track finding.
\end{enumerate}

\begin{figure}[!htb]
	\centering
	\includegraphics[width=0.95\columnwidth]{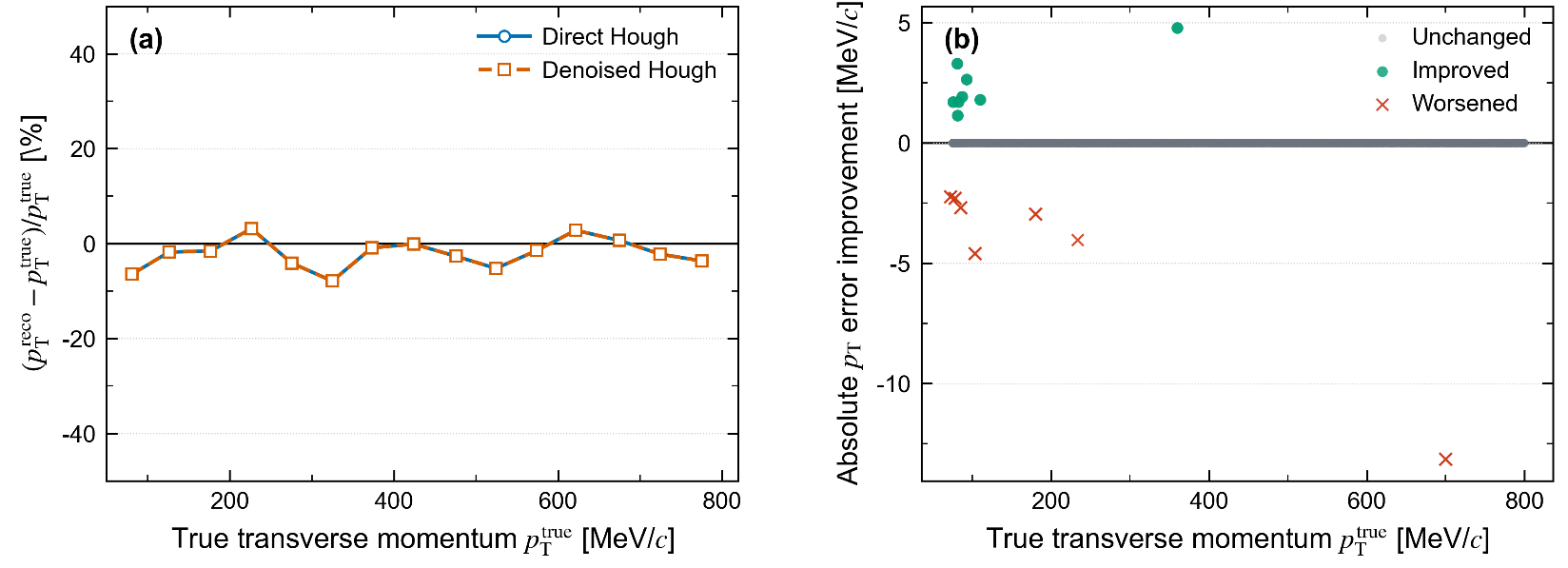}
	\caption{Comparison of the reconstructed transverse-momentum distributions for 5000 events processed using the direct Hough-transform method (Direct Hough) and the Hough-transform method after denoising (Denoised Hough). The true transverse-momentum range of the samples is $70$--$800~\mathrm{MeV}/c$. Among the 5000 events, only 15 events exhibit a change in the reconstructed transverse momentum caused by denoising, while the reconstruction results for all other events remain unchanged.}
	\label{fig:direct_vs_denoised_hough}
\end{figure}

Figure~\ref{fig:direct_vs_denoised_hough} shows the effect of transverse momentum when single-particle events with one nominal background level are reconstructed using the Hough-transform algorithm with and without denoising. For the vast majority of events, denoising does not interfere with the subsequent reconstruction process. Therefore, the denoising algorithm has almost no noticeable effect on the two-dimensional track-finding efficiency of the Hough-transform algorithm.

\begin{figure}[!htb]
	\centering
	\includegraphics[width=0.95\columnwidth]{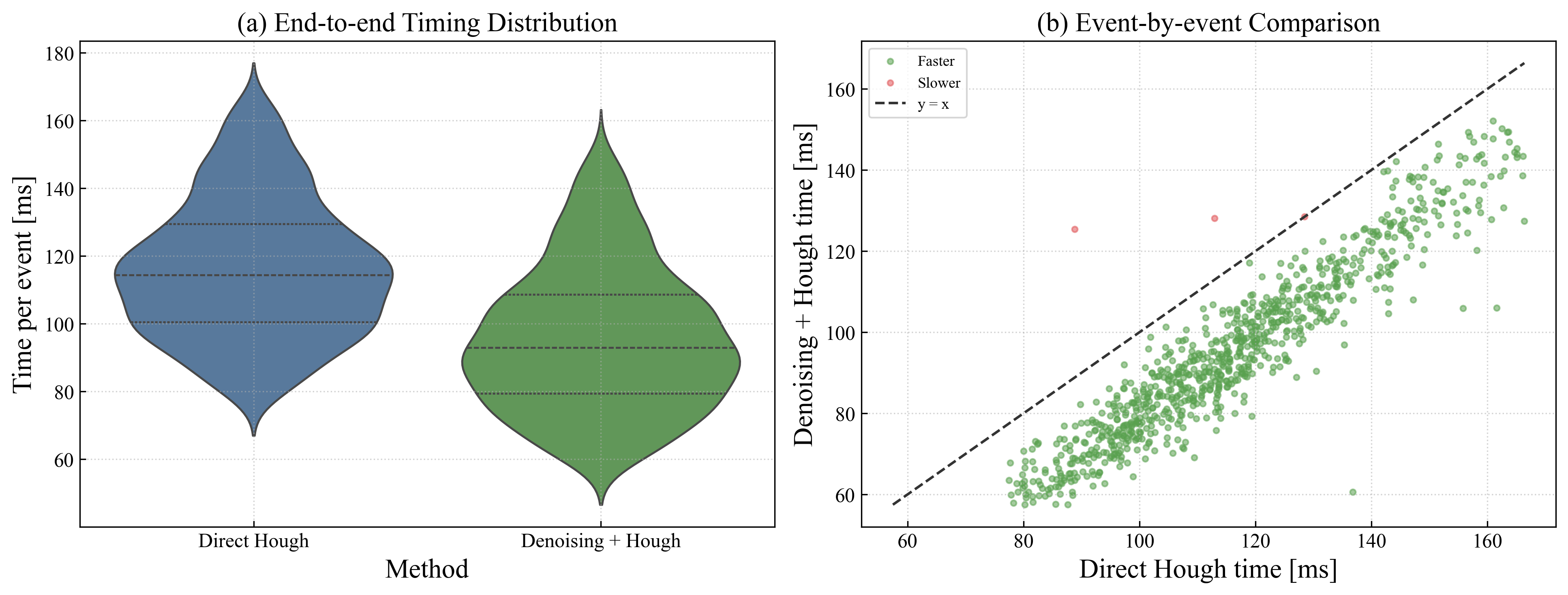}
	\caption{Comparison of the per-event processing time between direct two-dimensional Hough-transform track finding and two-dimensional Hough-transform track finding after denoising. The test samples contain 1000 single events with one nominal background level mixed into each event. The left panel shows the per-event processing-time distributions for the two processing workflows, and the right panel shows the per-event time comparison. By reducing the number of input signal points, denoising lowers the computational burden of Hough accumulation and peak finding. The average processing time per event decreases from approximately $116~\mathrm{ms}$ to approximately $96~\mathrm{ms}$, corresponding to an acceleration factor of approximately $1.21$. For a small number of events, denoising may modify the peak structure in Hough space, increase the number of candidate peaks, or make peak selection more complex. Consequently, the total processing time after denoising may be higher than that of the direct Hough-transform workflow for individual events.}
	\label{fig:hough_timing_comparison}
\end{figure}

 Figure~\ref{fig:hough_timing_comparison} compares the processing time of direct reconstruction using the Hough transform with that of reconstruction after denoising. Because the processing time of the Hough transform depends on the number of input hits, denoising reduces the overall processing time.

\subsection{Runtime Performance and Evaluation}

In the current design of the STCF HLT, the HLT is a heterogeneous computing cluster composed of CPUs and GPUs. For computations with sufficient parallelism, we aim to exploit the GPU for large-scale parallel processing in order to take advantage of its computational capability and energy efficiency. Accordingly, the current design aims to execute the early two-dimensional track-finding and reconstruction algorithms in the workflow on GPUs whenever possible. The denoising algorithm is therefore currently optimized and benchmarked mainly on GPUs.

We measured both the theoretical performance of the algorithm on the GPU and the practical performance and latency of the complete workflow when the CPU is used for data adaptation.

Modern GPUs mainly adopt the SIMT (Single Instruction, Multiple Threads) execution model~\cite{SIMT,CUDA}, in which one instruction is executed simultaneously by multiple threads. For data-independent models with substantial parallelism, GPUs generally provide higher parallelism and higher memory bandwidth than CPUs. At the same time, GPUs are less suitable for data flows with strong serial dependencies and complex branching. To improve hardware utilization and hide memory-access latency, each computation must provide many more threads than the number of available computing units. Therefore, in the GPU execution model, it is inefficient to transfer each event to the GPU and process it individually. Instead, hundreds of events are grouped into a batch and processed together by the GPU.

For parallel data organization, each sense wire is represented by one element in the two-dimensional matrix. The element type is \texttt{uint8}, with values ranging from 0 to 255. Each event consists of eight superlayer matrices, each with dimensions $6\times N$, where $N$ is determined by the superlayer. In memory, the complete batch of events is flattened into a contiguous one-dimensional array to reduce irregular memory access and simplify batch transfer and kernel scheduling.

The first step is sliding-window accumulation. Because the operations in this step are independent, multiple sliding windows can be processed in parallel. Each column of a superlayer matrix is assigned to a unique thread for updating, which avoids concurrent writes to the same element. Therefore, atomic operations are not required to ensure deterministic results. Shared intermediate quantities, including column masks, window-activation flags, and coverage counts, are used to reduce repeated calculations and maintain deterministic writes during parallel execution.

The second step is $3\times3$ local-neighborhood enhancement. In-place updating would cause newly computed values to participate in subsequent neighborhood calculations, introducing a cascade-propagation error. A separate buffer is therefore used. Each output position is still generated by a unique thread, so no write conflict occurs.

The third step is threshold filtering. The enhanced matrix is binarized according to the threshold: values greater than or equal to the threshold are set to one, and the remaining values are set to zero. This step is a regular element-wise transformation and is naturally suitable for parallel execution. In the implementation, threads are assigned by column, and the fixed six rows are traversed within each thread before writing the results back.

For the GPU thread organization, we use a one-dimensional grid and one-dimensional blocks. Each block corresponds to one superlayer matrix from one event. Threads within a block process columns in parallel, while the fixed row dimension of six is expanded within each thread. The grid size for a batch is $\mathrm{batch\_size}\times 8$, and the block size is fixed, for example at 256 threads, while remaining configurable. This thread model converts the explicit loops in the algorithm into implicit iterations over the thread space. The ``one unique write thread per column'' design avoids atomic conflicts.

\begin{figure}[!htb]
	\centering
	\includegraphics[width=0.95\columnwidth]{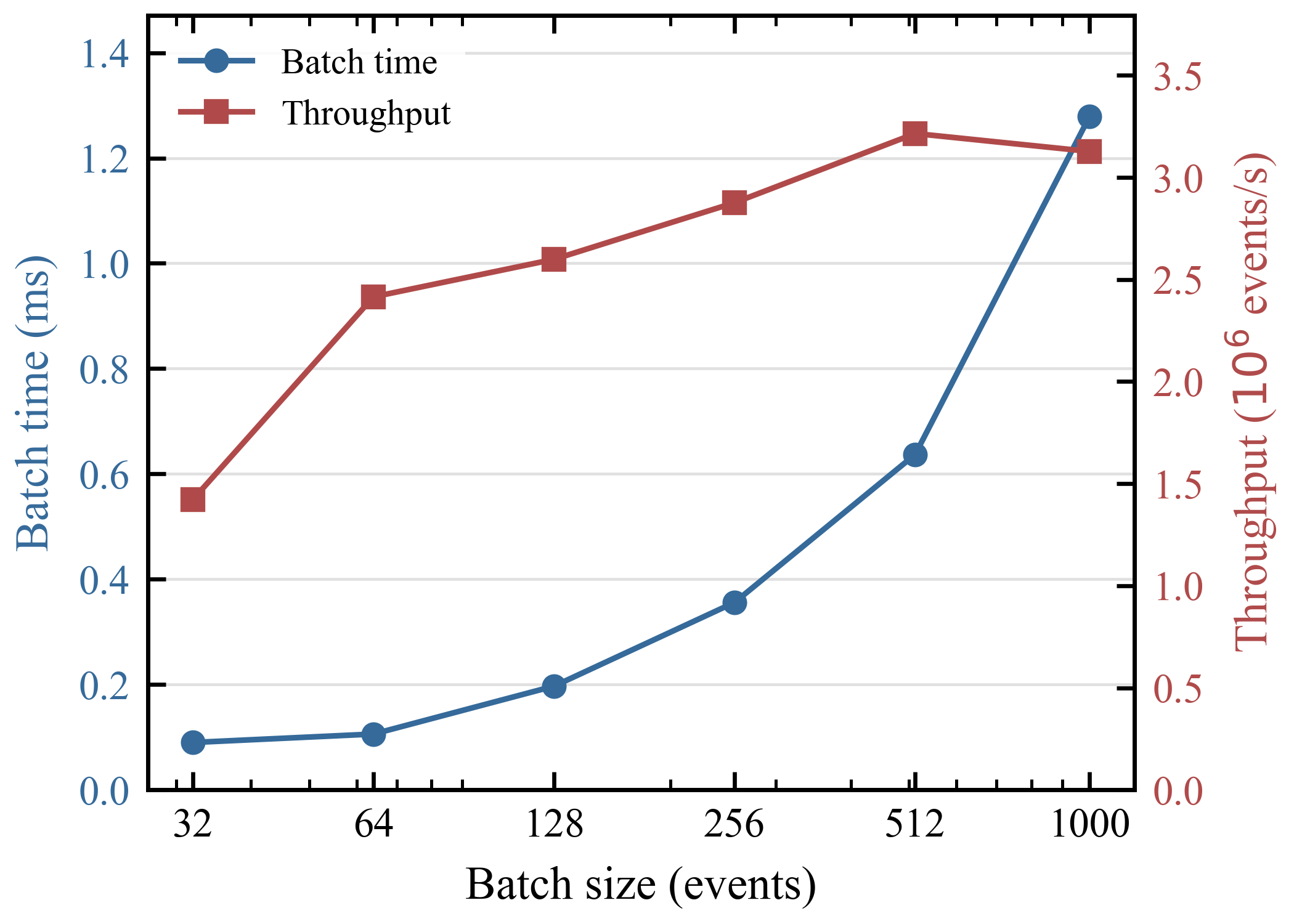}
	\caption{Batch-processing time and estimated throughput as functions of batch size. The throughput is calculated from the measured batch time and is multiplied by four to account for the four processing cores of one Tesla A16 GPU.}
	\label{fig:batch_throughput}
\end{figure}

Figure~\ref{fig:batch_throughput} illustrates the theoretical throughput and latency at different batch sizes using the Tesla A16 as an example. When the batch size is set to 128 events, the algorithm reaches a throughput of approximately $2.7\times10^{6}$ events/s on a single GPU of the Tesla A16. As the batch size increases, the per-event data footprint exceeds the cache capacity, causing a partial performance decrease.

We used NVIDIA Nsight Compute (NCU) to analyze the microarchitectural requirements of the key GPU resources. The results are listed in Table~\ref{tab:gpu_stage_perf}.

\begin{table}[t]
	\centering
	\caption{Utilization of key GPU resources at each processing stage, profiled using Nsight Compute.}
	\label{tab:gpu_stage_perf}
	\footnotesize
	\setlength{\tabcolsep}{2.5pt}
	\renewcommand{\arraystretch}{1.08}
	\begin{tabular}{lccc}
		\toprule
		Metric
		& \shortstack{\textbf{Stage 1}\\
			sliding-window\\[-1pt] accumulation}
		& \shortstack{\textbf{Stage 2}\\
			local-neighborhood\\[-1pt] enhancement}
		& \shortstack{\textbf{Stage 3}\\
			threshold\\[-1pt] filtering} \\
		\midrule
		\shortstack[l]{Average time\\($\mu\mathrm{s}$, full batch)}
		& 56.51 & 75.06 & 57.25 \\
		\shortstack[l]{Stage fraction (\%)}
		& 29.9 & 39.7 & 30.3 \\
		\shortstack[l]{Registers per thread}
		& 24 & 40 & 22 \\
		\shortstack[l]{Achieved occupancy}
		& 85.99\% & 75.73\% & 75.63\% \\
		\shortstack[l]{Compute (SM)\\throughput}
		& 81.50\% & 84.76\% & 31.08\% \\
		\shortstack[l]{L1/TEX cache\\throughput}
		& 84.55\% & 87.75\% & 28.66\% \\
		\shortstack[l]{DRAM throughput}
		& 29.52\% & 27.17\% & 31.46\% \\
		\shortstack[l]{Branch efficiency}
		& 100\% & 65.71\% & 100\% \\
		\bottomrule
	\end{tabular}
\end{table}

Under a batch size of 128, the GPU-computation part of the algorithm is stable and exhibits clear parallel advantages. Each batch requires approximately $188~\mu\mathrm{s}$, corresponding to an average processing time of approximately $1.46~\mu\mathrm{s}$ per event on one GPU. The three stages contribute approximately $29.9\%$, $39.7\%$, and $30.3\%$ of the processing time, respectively: the first stage, $6\times8$ sliding-window accumulation; the second stage, $3\times3$ local enhancement; and the third stage, threshold filtering. With a double-buffer implementation, each event occupies approximately $25~\mathrm{kB}$ of GPU memory.

Considering SM throughput, L1/TEX throughput, DRAM throughput, and occupancy together, the current program is not limited by a single pure-compute or pure-bandwidth bottleneck. Instead, it exhibits a mixed bottleneck caused by the combined effects of computation, on-chip-cache behavior, and control-flow overhead. This indicates that the current parallelization strategy is effective. Under the constraint of preserving the physics semantics, the loads of the three stages are relatively balanced, and no single subsystem is fully saturated in a way that would cause a severe mismatch.

In principle, the GPU could also construct the required data structures directly from the raw input. However, this approach would introduce considerable overhead. In a computer system, extracting or adapting data requires the original data to be loaded into memory first. This means that raw data not used during the HLT would still need to be transferred through the PCIe bus and occupy valuable GPU memory. In a streaming high-throughput environment, it is more efficient to discard irrelevant data as early as possible in the processing pipeline. Considering the cost difference between host memory and GPU memory, performing the data adaptation in host memory is currently more appropriate. This allows the GPU resources to be reserved for compute-intensive algorithms and makes better use of the GPU computational capability.

In the current R\&D phase, we mainly use the CPU for data adaptation. In a later phase, we plan to use an FPGA or a smart network interface card for this task. To ensure consistency, timing starts when the raw data are read into memory and stops when the data have been decomposed and reassembled into a two-dimensional \texttt{uint8} array. This interval is used to calculate the data-adaptation time. We read and processed $10\,\mathrm{k}$ events sequentially and used the average time as the processing time for one event.

The test environment was as follows: GCC 13.2 with the \texttt{-O2} optimization option, a Zen 4 architecture, one CPU core and one thread, and a fixed frequency of $2.5~\mathrm{GHz}$. Tests show that, on the HLT computing node, bit rearrangement and data loading for one event using a general-purpose CPU core require approximately $5.2~\mu\mathrm{s}$ on average. Therefore, for a batch size of 128, the latency of the denoising workflow is approximately $665~\mu\mathrm{s}$ for CPU assembly, $300~\mu\mathrm{s}$ for PCIe transfer, and $188~\mu\mathrm{s}$ for GPU processing, giving a total latency of approximately $1153~\mu\mathrm{s}$.

At a maximum event rate of $400~\mathrm{kHz}$, with all processing performed on the GPU, only approximately $14.6\%$ of one GPU's computational capacity is required, corresponding to approximately $0.3\%$ of the total GPU computational capacity of the complete system. In addition, because the data are greatly compressed after preprocessing, the bandwidth required to transfer the two-dimensional matrix data is approximately $10~\mathrm{GB/s}$ at the maximum event rate of $400~\mathrm{kHz}$. This is relatively low for the high-speed network and PCIe bus of the cluster. The one-way bandwidth of PCIe 4.0 $\times16$ is approximately $32~\mathrm{GB/s}$, and the actual deployment is expected to upgrade to PCIe 5.0. Therefore, the denoising algorithm is not expected to cause a congestion risk.

\section{Conclusion}

Drift chambers in high-luminosity particle collider experiments must operate continuously under high event rates and complex background conditions. The increasing volume of raw hit data places substantial pressure on online track finding, track reconstruction, and data transmission. To address this issue, this paper investigates an adaptive matrix-based denoising algorithm using local spatial support. The algorithm does not depend on complete track parameters, cross-superlayer pattern recognition, or global track fitting. An occupancy-dependent adaptive-threshold selection rule is obtained through offline calibration, improving the adaptability to different background conditions while maintaining low online computational overhead. Compared with fixed-threshold methods or conventional approaches that rely on manually tuned parameters, the event-level adaptive-threshold mechanism automatically adjusts the filtering parameters according to the occupancy and local aggregation degree of the current event, providing adaptive characteristics similar to those of a lightweight learning model. Compared with neural-network methods, the proposed algorithm requires neither training nor online inference in the neural-network sense. Its computation is simple, its resource consumption and processing latency are easy to control, and its decision criteria are relatively transparent and interpretable. Because the three main processing steps have regular data-access patterns and a high degree of data parallelism, the algorithm is well suited to batch-parallel processing on GPUs.

The proposed algorithm was validated in the HLT processing scenario of the STCF MDC, including both its physics performance and online deployment performance. The experimental results show that, under different physics channels and background levels, the algorithm removes $12.88\%$--$25.74\%$ of the input hits, with an average removal fraction of approximately $18.81\%$. The point-cluster background rejection rate ranges from $79.49\%$ to $96.67\%$, while the true-hit retention ratio remains above $98.48\%$. When the detector efficiency decreases to $90\%$, the true-hit retention ratio remains at 98\%, demonstrating that the algorithm is robust against local missing portions of genuine tracks. GPU tests show that the algorithm has processing latency on the order of $\mu\mathrm{s}$ and a high batch-processing throughput. On a Tesla A16, it requires approximately 14.6\% of the computational resources to satisfy the online processing requirements of the STCF MDC at an event rate of 400~$\mathrm{kHz}$. These results demonstrate that the algorithm achieves a favorable balance among genuine-hit retention, background suppression, and online computational overhead, and has potential for application as an upstream denoising algorithm in the STCF MDC HLT.

The research presented in this paper is not limited to the specific experimental conditions of the STCF. The core criterion of the algorithm is derived from the local spatial relationships among drift-chamber hits. Its implementation mainly relies on the geometric arrangement of the detector cells and the matrix representation of the input data, rather than on a specific physics process or predefined track parameters. Because multi-wire drift chambers in different experiments generally have layered and regularly arranged detector structures, a corresponding data mapping can be established according to the geometry of the target detector, and the window and threshold parameters can be recalibrated using simulation samples or calibration data. Therefore, the proposed algorithm has potential for transfer and extension to other drift chambers with similar regular structures. Future work will investigate the incorporation of event start times and drift-time information to improve the suppression of time-dependent backgrounds while maintaining a high true-hit retention ratio. The potential of this algorithm for addressing similar online hit-filtering problems in other high-luminosity detectors with regular geometries will also be explored.

\section{Acknowledgments}

This work is supported by the National Natural Science Foundation of China (No. 12341503); the Hunan Provincial Natural Science Foundation (No. 2023RC4006); the Scientific Research Fund of Hunan Provincial Education Department (No. 24B0454); the National Key R\&D Program of China (Nos. 2022YFA1602200 and 2023YFA1607200); and the international partnership program of the Chinese Academy of Sciences (No. 211134KYSB20200057). We thank the Hefei Comprehensive National Science Center for its support of the STCF key technology research project.

\end{document}